\documentclass[%
 aip,
cp,  
 amsmath,amssymb,
 reprint,%
]{revtex4-2}

\usepackage{graphicx}
\usepackage{dcolumn}
\usepackage{bm}

\usepackage[utf8]{inputenc}
\usepackage[T1]{fontenc}
\usepackage{mathptmx}

\usepackage[usenames,dvipsnames]{xcolor}
\usepackage{tcolorbox}
\usepackage{tabularx}
\usepackage{array}
\usepackage{colortbl}
\usepackage{multirow}

\tcbuselibrary{skins}

\newcolumntype{Y}{>{\raggedleft\arraybackslash}X}

\tcbset{tab1/.style={fonttitle=\bfseries\large,fontupper=\normalsize\sffamily,
colback=yellow!10!white,colframe=red!75!black,colbacktitle=Salmon!40!white,
coltitle=black,center title,freelance,frame code={
\foreach \n in {north east,north west,south east,south west}
{\path [fill=red!75!black] (interior.\n) circle (3mm); };},}}

\tcbset{tab2/.style={enhanced,fonttitle=\bfseries,fontupper=\normalsize\sffamily,
colback=yellow!10!white,colframe=red!50!black,colbacktitle=Salmon!40!white,
coltitle=black,center title}}

\newcommand{\newc}{\newcommand*}
\newc{\figurewidth}{3.2in}
\newc{\la}{\lambda}

\def\({\left(}
\def\){\right)}
\def\[{\left[}
\def\]{\right]}

\def\ee{\begin{equation}}
\def\q{\end{equation}}
\def\m{\begin{eqnarray}}
\def\n{\end{eqnarray}}
\def\a{\begin{aligned}}
\def\b{\end{aligned}}

\newc{\red}[1]{\textcolor{red}{#1}}
\newc{\yellow}[1]{\textcolor{yellow}{#1}}
\newc{\green}[1]{\textcolor{green}{#1}}
\newc{\blue}[1]{\textcolor{blue}{#1}}

\begin{document}

\title{Gravitational and electromagnetic radiations from binary black holes with electric and magnetic charges}

\author{Lang Liu} 
 \email[Corresponding author: ]{liulang915871@gmail.com}
 \affiliation{
  Department of Physics, Kunsan National University, Kunsan 54150, Korea.
}
\author{Sang Pyo Kim}%
 \email{sangkim@kunsan.ac.kr}
\affiliation{
  Department of Physics, Kunsan National University, Kunsan 54150, Korea.
}

\date{\today} 

\begin{abstract}
The Einstein-Maxwell theory has black hole solutions with electric and magnetic charges. In the standard model for particle physics, dyons with electric and magnetic charges would have been formed in the early universe.
 We derive the equations of motion of black hole binaries with electric and magnetic charges and explore some features of static orbits.  We calculate the total emission rates of energy and angular momentum due to gravitational and electromagnetic radiations from dyonic binary black holes in different cases.  It is shown that the emission rates of energy and angular momentum due to gravitational and electromagnetic radiations have the same dependence on the conic angle for different orbits. Moreover, we obtain the evolutions of orbits and find that a circular orbit remains circular while an elliptic orbit becomes quasi-circular due to electromagnetic and gravitational radiations. Our results provide rich information about black hole binaries with electric and magnetic charges and can be used to test black holes with magnetic charges.
\end{abstract}

\maketitle

\section{Introduction}

The Einstein-Maxwell theory has black hole solutions with electric and/or magnetic charges, known as the Reissner-Nordstr\"{o}m black holes or Kerr-Newman black holes in the asymptotically flat space or (anti-) de Sitter space. It is highly probable that magnetic monopoles and dyons both with electric and magnetic charges predicted by the standard model of particle physics would have produced, and it is also likely that charged black holes might have been formed in the early universe. Those exotic particles or black holes would encounter to form binaries of pairs of black holes with the same or opposite charges.

Magnetic charges, if they exist in the early Universe, will provide a hitherto unexplored window to probe fundamental physics in the Standard Model of particle physics and beyond. Though no evidence of magnetic charges has been found yet ~\cite{Staelens:2019gzt,2020arXiv200505100M}, magnetically charged black holes have attracted much attention not only in theoretical study but also in recent astronomical observations \cite{Maldacena:2020skw, Bai:2020spd, Ghosh:2020tdu}. Recently, restoration of the electroweak symmetry near the horizon of a magnetic charged black hole has been discussed in \cite{Maldacena:2020skw} and the phenomenology of low-mass magnetic black holes, which can have electroweak-symmetric coronas outside of the event horizon, has been comprehensively studied in \cite{Bai:2020spd}. Potential astrophysical signatures for magnetically charged black holes have also been investigated in \cite{Ghosh:2020tdu}.

A dyonic black hole is a {nonrotating or rotating black hole with an electric charge $q$ and a magnetic charge $g$. A dyonic nonrotating black hole} has the same metric as the Reissner-Nordstr\"{o}m black hole with $q^2$ replaced by $q^2+g^2$ \cite{1982PhRvD..25..995K}. In the Minkowski spacetime, the nonrelativistic interaction of two dyons was studied in a quantum theory~\cite{1968PhRv..176.1480Z} and in classical theory~\cite{1976AnPhy.101..451S}.

According to the ``no-hair" conjecture, a general relativistic black hole is completely described by four parameters: mass, angular momentum, magnetic charge as well as electric charge. Compared to Schwarzschild black holes, charged black holes have rich phenomena. Recently, charged black holes have been discussed extensively \cite{Cardoso:2016olt,Liebling:2016orx,Toshmatov:2018tyo,Bai:2019zcd,Allahyari:2019jqz,Christiansen:2020pnv,Wang:2020fra,Bozzola:2020mjx,Kim:2020bhg,Cardoso:2020nst,Cardoso:2020iji, McInnes:2020gxx,Bai:2020ezy,Bozzola:2021elc,McInnes:2021frb,Hou:2021suj,Benavides-Gallego:2021the}. Binary black holes with charges emit not only gravitational radiation but also electromagnetic radiation.

In the Newton theory, the central forces of gravitation and Coulomb interaction confine the orbit of the electrically charged object on a plane, and a pair of masses and charges form a binary when the Coulomb repulsion does not exceed the gravitational attraction. This means that charged black holes can form a binary when the combined gravitational and electric interactions of is attractive. The binary follows a circular or elliptical orbit and radiates both the gravitational and electromagnetic waves. 

On the other hand, a binary of dyonic black holes both with electric and magnetic charges experiences an additional force from the velocity-dependent magnetic force and has a conserved angular momentum different from the orbital angular momentum. The binary precesses with respect to the conserved angular momentum and follows circular orbits or elliptical orbits with a finite or infinite winding number on a Poincare cone. Those orbits with different winding numbers on the cone exhibit a rich structure of orbital motion.

In this paper, we review the recent works on gravitational and electromagnetic radiations from binary black holes with electric and/or magnetic charges and the evolution of the binary due to the loss of binding energy and angular momentum. This paper is organized as follows: (i) the orbits under the gravitational attraction and the electromagnetic interaction due to charges, (ii) gravitational and electromagnetic radiations in the inspiral motion, and (iii) the evolution of orbits due to the radiations. For more details, please refer to \cite{Liu:2020vsy,Liu:2020bag}. Throughout this paper, we set $G=c =4 \pi \varepsilon_{0} = \mu_0 /4\pi =1$.

\section{\label{II}Orbits without radiation}

The setup of this paper is a binary of black holes with electric and magnetic charges, that is, dyonic black holes. We shall consider the weak-field configuration, where the semimajor axis of the dyonic black hole binary is much larger than their event horizons and the backreaction of the binary on the spacetime geometry can be safely neglected. In such a case, a dyonic black hole binary during the inspiral motion is well approximated by a pair of massive point-like objects with electric and magnetic charges.

Using a Newtonian method, we study the orbital motion of a binary of black holes with electric and magnetic charges. Maxwell's equations with magnetic monopoles are given by
\m
\boldsymbol{\nabla} \cdot \boldsymbol{E}  &=& 4 \pi \rho_{\mathrm{e}}, \quad \boldsymbol{\nabla} \cdot \boldsymbol{B}  =4\pi \rho_{\mathrm{m}},  \nonumber\\
\boldsymbol{\nabla} \times \boldsymbol{E} & =& -4\pi \boldsymbol{j}_{\mathrm{m}}-\partial \boldsymbol{B} / \partial t , \quad
\boldsymbol{\nabla} \times \boldsymbol{B}  = 4\pi \boldsymbol{j}_{\mathrm{e}}+\partial \boldsymbol{E} / \partial t.
\n
where $\rho_{\mathrm{e}}$ and $\rho_{\mathrm{m}}$ are electric and a magnetic charge densities and $\boldsymbol{j}_{\mathrm{e}}$ and $\boldsymbol{j}_{\mathrm{m}}$ are their electric and magnetic currents, respectively.  The Lorentz force on a dyon with an electric charge $q$ and a magnetic charge $g$ is
\m
\mathbf{F}= q(\mathbf{E}+\mathbf{v} \times \mathbf{B})+
g\left(\mathbf{B}-\mathbf{v} \times \mathbf{E}\right).
\n
A point dyon generates the electric and magnetic fields:
\m
\mathbf{E} = q \frac{\mathbf{r}}{r^3}, \quad \mathbf{B} = g \frac{\mathbf{r}}{r^3}.
\n

\begin{figure}[htpb]
    \includegraphics[width=0.48\textwidth]{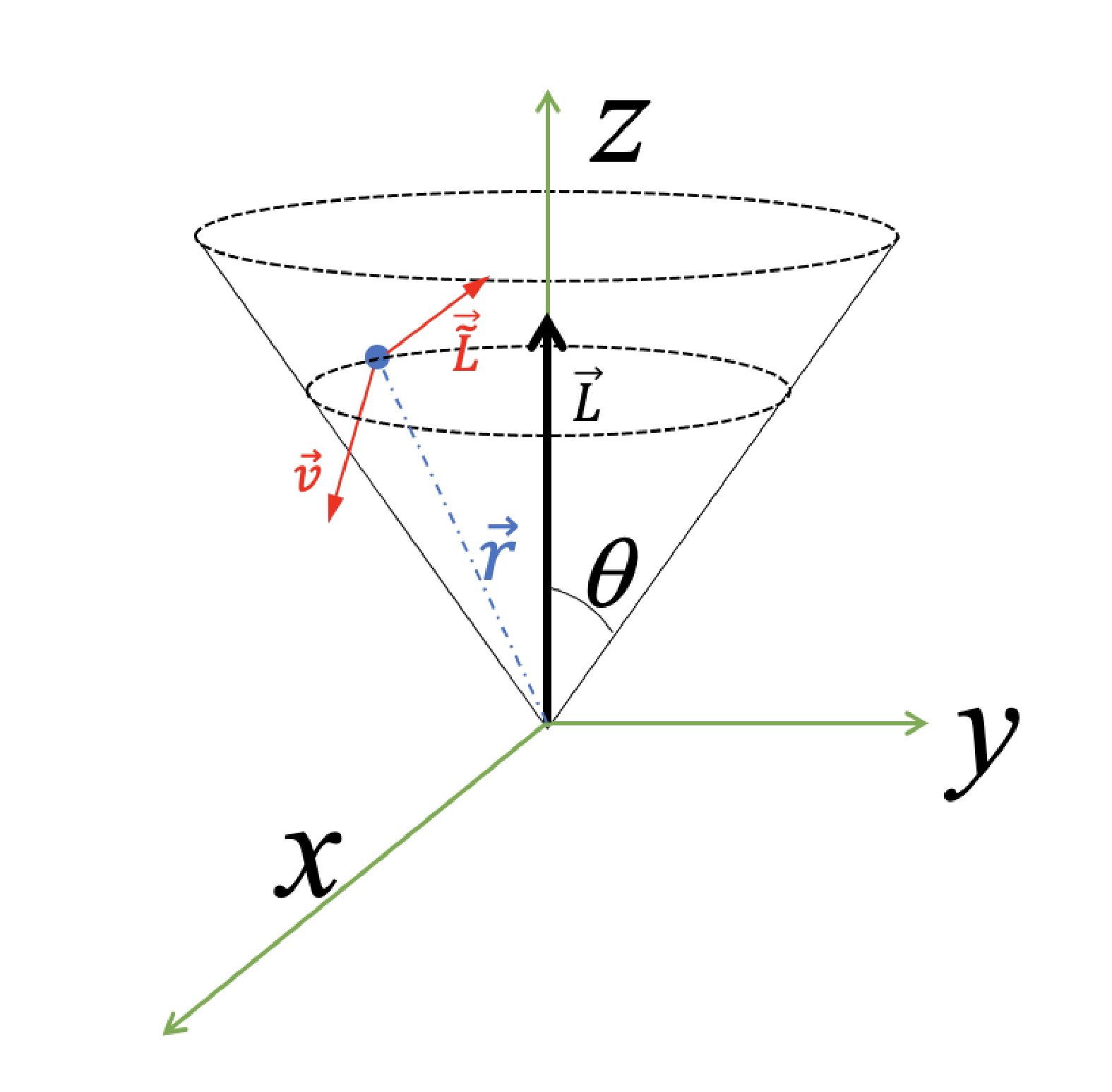}
    \caption[]{The plot of a Poincare cone with the $z$ axis along the generalized angular momentum. The binary orbit is instantaneously perpendicular to $\mathbf{\tilde{L}}$.}
   \label{fig:cone}
\end{figure}

For a binary of  dyonic black holes, we choose the origin at the center of mass system:
\m
{\mathbf r}_{1} = -\frac{m_2}{M} {\mathbf r}, \quad {\mathbf r}_{2} = \frac{m_1}{M}{\mathbf r},
\n
where
\m
{\mathbf r} = {\mathbf r}_{2} - {\mathbf r}_{1}, \quad M = m_1+m_2.
\n
The equation of motion under the Lorentz force and gravitational force is given by
\m
\label{eom}
\mu \ddot{\mathbf r} = C \frac{{\mathbf r}}{r^{3}}- D \frac{{\mathbf r} \times {\mathbf v}}{r^{3}},
\n
where
\m
\mu=\frac{m_1m_2}{M}, \quad {\mathbf v}=\dot{{\mathbf r}},
\n
and
\m \label{C-D}
C=\left(-\mu M+q_{1} q_{2}+g_{1} g_{2}\right), \quad  D=\left(q_{2} g_{1}-g_{2} q_{1}\right).
\n
It is not the orbital angular momentum  $\mathbf{\tilde{L}} \equiv \mu \mathbf{r} \times \mathbf{v}$ but the generalized angular momentum, known as the Laplace-Runge-Lenz vector, of binary system $\mathbf{L}$ defined by $\mathbf{L}\equiv \mathbf{\tilde{L}}-D \mathbf{r}/r$ that is conserved.
Choosing the $z$-axis along $\mathbf{L}$ as shown in Fig.~\ref{fig:cone}, the orbit is given by
\m
\label{R}
{\mathbf r} = \frac{a\left(1-e^{2}\right)}{1+e \cos (\phi \sin \theta)}\left(\begin{array}{c}
\sin \theta \cos \phi \\
\sin \theta \sin \phi \\
\cos \theta
\end{array}\right) ,
\n
where $a$ and $e$ can be interpreted as the semimajor axis and eccentricity, which are defined by
\m
\label{E}
a \equiv \frac{C}{2E}, \quad e \equiv \left(1-\frac{2E \tilde{L}^{2}}{\mu C^2}\right)^{1/2}.
\n
Here, $E$ is the orbital energy of the binary with the parameter $C$ for the inverse square-law and $\theta=\arccos (|D| / L)$ is a constant due to the conservation of the generalized angular momentum. Throughout this work, we only consider $0<\theta\leq \pi/2$. (For $\pi/2\leq\theta< \pi$, we can redefine $\mathbf{r} = \mathbf{r}_{1} - \mathbf{r}_{2}$ to make $0<\theta\leq \pi/2$.) In Eq.~(\ref{R}) we have expressed the quantities in the denominator and numerator with the conic section parameters $(a, e)$ so that the orbit has a standard form on the cone. {The $\vec{\tilde{L}}$ is instantaneously perpendicular to the orbit, and the angle $\vartheta$ between $\mathbf{\tilde{L}}$ and $\mathbf{L}$ satisfies the relation $\vartheta + \theta = \pi/2$, which follows from the relations $L^2 = \tilde{L}^2 + D^2$ and $\cos \theta = |D|/L$. Hence, the instantaneous orbit precesses with a fixed angle $\vartheta$ around $\bm{L}$.}
From now on we shall consider only the case when the binary system is bounded, in other words, $E<0$, $(C<0)$. Thus, when the electromagnetic interaction is repulsive, it must be smaller than the gravitational attraction. From (\ref{R}) the binary orbits a closed loop when $\phi \sin \theta = 2 \pi l$ and $\phi = 2 \pi n$ for a pair of natural numbers $l$ and $n$, that is, $\sin \theta = l/n$. On the Poincare cone, the radius $r$ nutates $l$-times between the perihelion and aphelian and while $\phi$ revolves $n$-times. The evolution of the azimuthal angle is
\m
\label{dphidt}
\dot{\phi}=\frac{(-C)^{\frac{1}{2}} \csc (\theta ) (e \cos (\phi  \sin (\theta ))+1)^2}{a^{\frac{3}{2}} \left(1-e^2\right)^{\frac{3}{2}} \mu^{\frac{1}{2}} }.
\n

\begin{figure}[htpb]
    \includegraphics[width=0.48\textwidth]{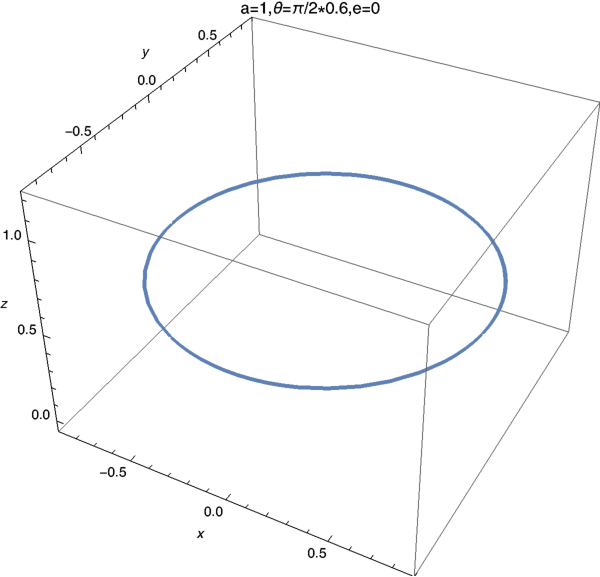}
    \caption[]{For $ e=0$, $x^2 + y^2 = a^2 \sin^2 \theta$ and $z = a \cos \theta$, which describes a circle in the $x-y$ plane, regardless of the value $\theta$. A circle-shaped orbit of the binary is plotted for the parameters $a=1,\, \theta = 3 \pi / 10$.  Though $\sin(3 \pi/10) = (1+ \sqrt{5})/4$ is an irrational number, the orbit is closed and has a finite period in three dimensions~\cite{Liu:2020vsy}.}
   \label{fig:e=0}
\end{figure}

Now that we have orbits of dyonic binary, we will explore the features in general.
First, let us consider the case $e=0$ for our binary system. From Eqs.~\eqref{dphidt} and \eqref{R}, the three-dimensional trajectory
is effectively a two-dimensional circular orbit with $z = \cos \theta$, as illustrated in Fig.~\ref{fig:e=0}, and the orbital rate
\m
\label{dphidte=0}
\dot{\phi}=\frac{(-C)^{\frac{1}{2}}}{\mu^{\frac{1}{2}} a^{\frac{3}{2}} \sin \theta },
\n
has a finite period, as expected for Keplerian orbits,
\m \label{T-1}
 T_1=\int_0 ^{2\pi} d\phi \dot{\phi}^{-1}=2 \pi a^{3 / 2} \sqrt{-\mu / C} \sin \theta.
\n

\begin{figure}[htpb]
     \includegraphics[width=0.4\textwidth]{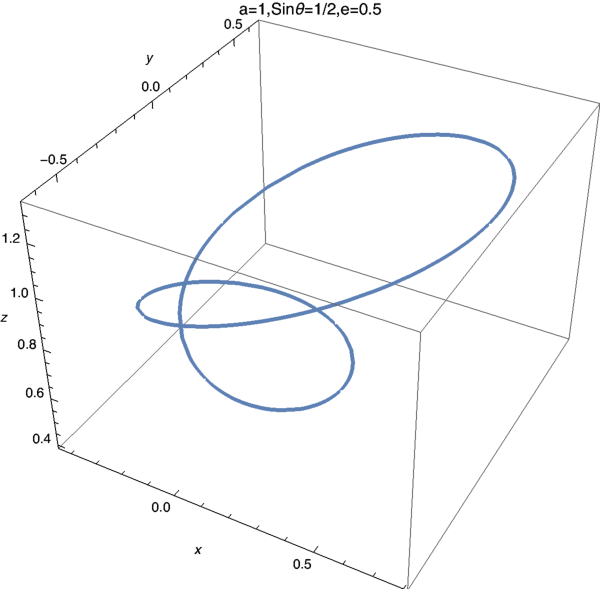}
     \includegraphics[width=0.4\textwidth]{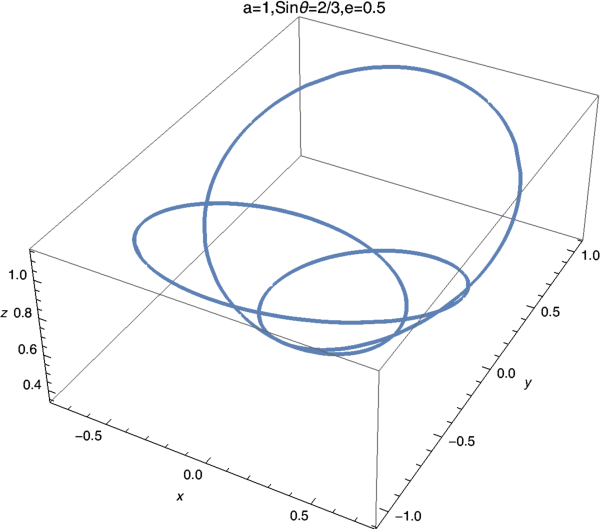}
    \caption[]{ Two different closed orbits of the binary with the parameters $a=1 $ and $e=0.5$ are illustrated for the rational values of  $\sin \theta=1/2$ (left) and $\sin \theta=2/3$ (right).}
    \label{fig:R4}
\end{figure}

Next, we consider a conical elliptical orbit of $e \neq 0$. In the case of a rational $\sin \theta$,
\m
\sin \theta=\frac{l}{n}
\n
with $l$ and $n$ relative prime numbers $(l<n)$, the orbit is closed after $n$ revolutions and the system will complete one exact ellipse and return to the initial position. In such a case, one period is given by
\m \label{T-2}
T_2=\int_0 ^{2n\pi} d\phi \dot{\phi}^{-1}=2 \pi a^{3 / 2} \sqrt{-\mu / C} l.
\n
Moreover, the different integers $l$ and $n$ give the different topology of the orbit, as shown in Fig.~\ref{fig:R4}.

\begin{figure}[htpb]
    \includegraphics[width=0.48\textwidth]{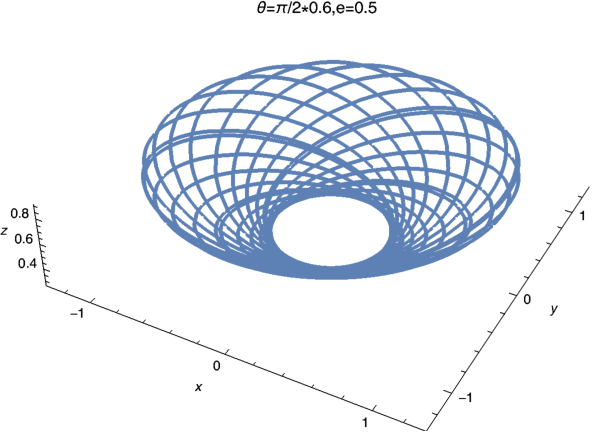}
    \caption[]{ A conic-shaped orbit of the binary which is confined to the surface of a cone is plotted in the range of $\phi$ from $0$ to $40 \pi/\sin \theta$ and the parameters $a=1,\, e=0.5$ and $\theta= 3\pi/10$ according to \eqref{R}.  Though the orbit is bounded, it is not closed and has an infinite period in three dimensions since $\sin(3 \pi/10)$ is an irrational number~\cite{Liu:2020vsy}.}
    \label{fig:R}
\end{figure}

\begin{figure}[htpb]
    \includegraphics[width=0.48\textwidth]{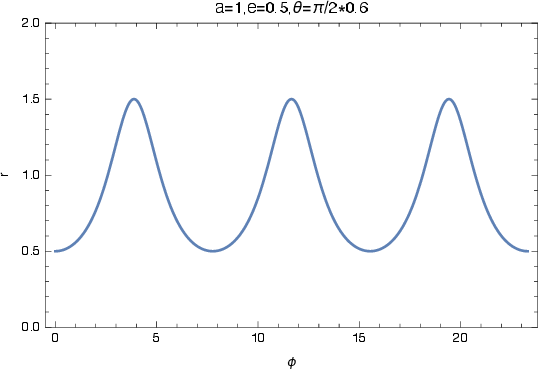}
    \caption[]{The plot of $r = a (1-e^2)/(1+ e \cos(\phi \sin \theta))$ as a function of $\phi$ by choosing $a=1, \theta= 3\pi/10$ and $e=0.5$~\cite{Liu:2020vsy}. It has a period of $2 \pi/\sin (3 \pi/10)$.}
    \label{fig:absR}
\end{figure}

When $e \neq 0$ and $\sin \theta$ is irrational, the orbit is not closed and shows a chaotic behavior of a conserved autonomous system \cite{Argyris:2015311}; for instance,  in Fig.~\ref{fig:R},  we plot the orbit of the binary by choosing $a=1, \theta= 3 \pi/10$ and $e=0.5$. For $e \neq 0$, no matter how rational or irrational $\sin \theta$ is, $r$ is a periodic function of $\phi(t)$ with the period
\m \label{T-3}
 T_3=\int_0 ^{2\pi/\sin(\theta)} d\phi \dot{\phi}^{-1}=2 \pi a^{3 / 2} \sqrt{-\mu / C},
\n
as shown in Fig.~\ref{fig:absR}.

\section{\label{III} Electromagnetic and gravitational radiations}
 Static orbits of dyonic black hole binaries {on a conic section are divided into three categories: (i) $e=0$, (ii) $e \neq 0$ and rational $\sin \theta$ and (iii) $e \neq 0$ and irrational $\sin \theta$. The orbits of dyonic black hole binaries in those different cases have different topologies.  In this section, we review the loss of energy and angular momentum due to gravitational and electromagnetic radiations for the three different cases in~\cite{Liu:2020vsy,Liu:2020bag}.  The averaged rate of, for instance, the momentum loss
 \m
 \left\langle \frac{d \mathbf{L}}{dt} \right\rangle \equiv
\lim _{ T \rightarrow \infty }  \frac{1}{T} \int_{0}^{T} dt \frac{d \mathbf{L}}{dt} = \lim _{ T \rightarrow \infty } \frac{1}{T} \int_{0}^{T \dot{\phi}} \frac{d \phi}{\dot{\phi}} \frac{\mathbf{L}}{dt}.
 \n
A similar expression holds for the rate of the energy loss. In the first case of $e = 0$, the $\phi$ has the period of $2 \pi$ and thus, the period for the one revolution is $T_1 = 2 \pi / \dot{\phi} $ in \eqref{T-1}. In the second case of $e \neq 0$ and $\sin \theta = l/n$, the orbits returns to the initial position after $\phi$ revolves $n$-times and $r$ nutates $l$-times, which give the period $T_2 =2 n \pi / \dot{\phi}$ in \eqref{T-2}. However, in the case of an irrational $\sin \theta$, the orbit does not close a loop, so we take $\phi = 2 N \pi /\sin \theta $ for arbitrary large $N$ and the period $T = N T_3$ with $T_3$ in \eqref{T-3}.

\subsection{\label{e=0} Case: $e=0$}
Following \cite{Liu:2020vsy}, the energy and angular momentum emission due to electromagnetic radiation are given by
 \m
 \frac{dE_{EM}}{dt}=-\frac{2 \mu^{2}((\Delta \sigma_q)^{2}+(\Delta \sigma_g)^{2})}{3} \ddot{r}^i \ddot{r}_i,
 \n
 \m
\frac{dL^i_{EM}}{dt}=-\frac{2 \mu^{2}((\Delta \sigma_q)^{2}+(\Delta \sigma_g)^{2})}{3} \epsilon_{j k}^{i}\dot{r}^{j} \ddot{r}^{k}
 \n
where $\epsilon^{i}_{j k}$ is the Levi-Civita symbol and
\m
\Delta \sigma_q=q_{2} / m_{2}-q_{1} / m_{1},\quad
\Delta \sigma_g=g_{2} / m_{2}-g_{1} / m_{1},
\n
are the dipole moments of electric charges and magnetic charges.
The averaged energy loss over an orbital period $T_1$ due to  electromagnetic radiation is given by
\m
\left\langle\frac{dE_{EM}}{dt}\right\rangle = -\frac{2 C^2 ((\Delta \sigma_q)^{2}+(\Delta \sigma_g)^{2}) \csc ^2(\theta )}{3 a^4}.
\n
The averaged angular momentum loss over an orbital period $T_1$ due to  electromagnetic radiation is
\m
\left\langle\frac{dL^i_{EM}}{dt}\right\rangle \equiv \frac{1}{T_1} \int_{0}^{T_1} d t \dot{L}^i_{EM}.
\n
After a straightforward computation, we obtain the averaged angular momentum loss
\m
\left\langle\dot{L}_{EM}^1\right\rangle=\left\langle\dot{L}_{EM}^2\right\rangle=0,
\n
\m
\left\langle\dot{L}_{EM}^3\right\rangle=-\frac{2  (-C)^{3/2} \sqrt{\mu} ((\Delta \sigma_q)^{2}+(\Delta \sigma_g)^{2}) \csc (\theta )}{3 a^{5/2}}.
\n

Now, we compute the total radiated power in gravitational waves.  In our reference frame where $\bm{L}$ is along the $z$ axis, the second mass moment can be written as
\m
\label{Mij}
M^{ij}=\mu r^i r^j.
\n
Following \cite{Peters:1963ux} and \cite{Peters:1964zz}, the energy and angular momentum emission due to gravitational radiation are given by
\m
\frac{dE_{GW}}{dt}=-\frac{1}{5 }\left\langle\ddot{M}_{i j} \ddot{M}_{i j}-\frac{1}{3}\left(\ddot{M}_{k k}\right)^{2}\right\rangle,
\n
\m
\frac{d L^{i}_{GW}}{d t}=-\frac{2}{5} \epsilon^{i k l}\left\langle\ddot{M}_{k a} \ddot{M}_{l a}\right\rangle.
\n
Similarly, we have the energy and angular momentum loss due to gravitational radiation averaged one orbital period $T_1$:
\m
\left\langle\frac{dE_{GW}}{dt}\right\rangle=\frac{(-C)^3 (15 \cos (2 \theta )-17) \csc ^4(\theta )}{5 a^5 \mu },
\n
\m
\left\langle\frac{dL_{GW}^{3}}{dt}\right\rangle=\frac{(-C)^{5/2} (15 \cos (2 \theta )-17) \csc ^3(\theta )}{5 a^{7/2} \sqrt{\mu }},
\n
and $\left\langle\dot{L}_{GW}^1\right\rangle=\left\langle\dot{L}_{GW}^2\right\rangle=0.$

\subsection{\label{rat sin} Case: $e\neq0$ and rational $\sin \theta$}

In this subsection, we calculate the emissions of energy and angular momentum from the orbit with $e \neq 0$ and rational $\sin \theta = l/n$ where $n$ and $l$ are relative prime numbers and $l<n$.

First, we calculate the emission of electromagnetic radiation due to the electric and magnetic charges on the orbit \eqref{R}, averaged over one orbit
\m
\left\langle\dot{L}_{EM}^1\right\rangle=\left\langle\dot{L}_{EM}^2\right\rangle=0,
\n
and
\ee
\label{A1}
\left\langle\frac{dE_{EM}}{dt}\right\rangle
= -\frac{C^2((\Delta \sigma_q)^{2}+(\Delta \sigma_g)^{2}) n^2}{24 a^4 \left(1-e^2\right)^{5/2} l^2}
\times \left(3 e^4+\left(3 e^2+20\right) e^2 (1-\frac{2 l^2}{n^2})+28 e^2+16\right),
\q
\ee
\label{A2}
\left\langle\dot{L}_{EM}^3\right\rangle=-\frac{  (-C)^{3/2} \sqrt{\mu} ((\Delta \sigma_q)^{2}+(\Delta \sigma_g)^{2}) n }{6 a^{5/2} (1-e^2) l}
\Bigl( e^2 (2-\frac{2 l^2}{n^2} )+4 \Bigr).
\q

Second, the averaged energy and angular momentum loss rates due to gravitational radiation are
\m
\label{A3}
&&\left\langle\frac{dE_{GW}}{dt}\right\rangle=\frac{(-C)^3 n^4}{240 a^5 \left(1-e^2\right)^{7/2} l^4 \mu }
\times \Bigl( 2 \left(e^2+1\right) \left(15 e^2+308\right) e^2 (\frac{8 l^4}{n^4}-\frac{8 l^2}{n^2}+1)
\notag \\
&+&\left(-15 e^6+26 e^4+1976 e^2+720\right) (1-\frac{2 l^2}{n^2})
-3 \left(15 e^6+404 e^4+1104 e^2+272\right) \Bigr),
\n

\ee
\label{A4}
\left\langle\dot{L}_{GW}^3\right\rangle=\frac{(-C)^{5/2 } n^3}{40 a^{7/2} \sqrt{\mu }(1-e^2)^2 l^3}  \Bigl( \left(7 e^2+48\right) e^2 (\frac{8 l^4}{n^4}-\frac{8 l^2}{n^2}+1)-2 \left(5 e^4+92 e^2+68\right)
\left(-3 e^4+88 e^2+120\right) (1-\frac{2 l^2}{n^2}) \Bigr),
\q
and $\left\langle\dot{L}_{GW}^1\right\rangle=\left\langle\dot{L}_{GW}^2\right\rangle=0$.

\subsection{\label{irr sin} Case: $e\neq 0$ and irrational $\sin \theta$}

The orbit does not close its trajectory, so the period is infinite, which we take $T= NT_3$ for large $N$ while the $\phi$ elapses $2 N \pi/\sin \theta$. Thus, the energy and angular momentum emission due to electromagnetic radiation leads to the averaged energy loss rate
\m
\left\langle\frac{dE_{EM}}{dt}\right\rangle=\lim _{ N\rightarrow \infty }\frac{1}{N T_3} \int_0 ^{2N\pi/\sin \theta} d\phi \frac{dE_{EM}}{dt} \dot{\phi}^{-1},
\n
\m
\left\langle\dot{L}^{i}_{EM}\right\rangle=\lim _{ N\rightarrow \infty }\frac{1}{N T_3} \int_0 ^{2N\pi/\sin \theta} d\phi \dot{L}^{i}_{EM} \dot{\phi}^{-1}.
\n
So, we get
\m
\label{B1}
&&\left\langle\frac{dE_{EM}}{dt}\right\rangle=- \frac{C^2((\Delta \sigma_q)^{2}+(\Delta \sigma_g)^{2}) \csc ^2(\theta )}{24 a^4 \left(1-e^2\right)^{5/2} } \Bigl(3 e^4+\left(3 e^2+20\right) e^2 \cos (2\theta)+28 e^2+16 \Bigr),
\n
\m
\label{B2}
\left\langle\dot{L}_{EM}^3\right\rangle&=&-\frac{  (-C)^{3/2} \sqrt{\mu} ((\Delta \sigma_q)^{2}+(\Delta \sigma_g)^{2}) \csc (\theta ) }{6 a^{5/2} (1-e^2) }
 \(e^2 \cos (2 \theta )+e^2+4\),
\n
and $\left\langle\dot{L}_{EM}^1\right\rangle=\left\langle\dot{L}_{EM}^2\right\rangle=0$.
Similarly, we find the averaged energy loss due to gravitational radiation is given by
\m
\label{B3}
&&\left\langle\frac{dE_{GW}}{dt}\right\rangle=\frac{(-C)^3 \csc ^4(\theta )}{240 a^5 \left(1-e^2\right)^{7/2}  \mu }
 \Bigl( 2 \left(e^2+1\right) \left(15 e^2+308\right) e^2 (\cos (4 \theta ))
\notag \\
&+&\left(-15 e^6+26 e^4+1976 e^2+720\right) (\cos (2 \theta ))
-3 \left(15 e^6+404 e^4+1104 e^2+272\right) \Bigr),
\n
and the averaged angular momentum loss rate by
\ee
\label{B4}
\left\langle\dot{L}_{GW}^3\right\rangle=\frac{(-C)^{5/2 } \csc ^3(\theta )}{40 a^{7/2} \sqrt{\mu }(1-e^2)^2 }
 \Bigl( \left(7 e^2+48\right) e^2 (\cos (4 \theta ))-2 \left(5 e^4+92 e^2+68\right)
+\left(-3 e^4+88 e^2+120\right) \cos (2 \theta ) \Bigr),
\q
while
\m
\left\langle\dot{L}_{GW}^1\right\rangle=\left\langle\dot{L}_{GW}^2\right\rangle=0.
\n

\section{Total Emission Rates of Energy and Angular Momentum}

In the above we have found, considering the periodicity, the averaged rates for energy and angular momentum loss due to gravitational and electromagnetic radiations for three different cases. The rates obtained by taking an infinite time average in the case of irrational $\sin \theta$ involve sinusoidal functions of $\theta$, and we check whether these are the most general formulae. By replacing $\sin \theta = l/n$ and using the relation
\m
\cos (2 \theta )=1-\frac{2 l^2}{n^2},  \quad
\cos (4 \theta )=\frac{8 l^4}{n^4}-\frac{8 l^2}{n^2}+1,
\n
one can show that \eqref{B1}, \eqref{B2}, \eqref{B3} and \eqref{B4} are equal to \eqref{A1}, \eqref{A2}, \eqref{A3} and \eqref{A4}. Further, note that \eqref{B1}, \eqref{B2}, \eqref{B3} and \eqref{B4} are also valid for the case $e=0$, which can be shown by $\sin \theta = \sqrt{-a C \mu}/\sqrt{-a C \mu +D^2}$. As expected from physical reason and intuition, the continuous loss of the energy and angular momentum due to gravitational and electromagnetic radiations decreases the semimajor axis $a$ and the eccentricity $e$, which in turn continuously decreases $\theta$. Therefore, it is consistent and legitimate to use the loss formulae in \eqref{B1}, \eqref{B2}, \eqref{B3} and \eqref{B4} for all $\sin \theta$, $a$ and $e$.

Now that the emission rates of energy and angular momentum \eqref{B1}, \eqref{B2}, \eqref{B3} and \eqref{B4} are valid for all $\sin \theta$, $a$ and $e$, the total emission rates of energy and angular momentum are given by
\m
\label{dEtotal}
\left\langle\frac{dE}{dt}\right\rangle &=& \left\langle\frac{dE_{EM}}{dt}\right\rangle+\left\langle\frac{dE_{GW}}{dt}\right\rangle \nonumber\\
&=& \frac{((\Delta \sigma_q)^{2}+(\Delta \sigma_g)^{2}) (-C) \left(4 a \left(e^4+e^2-2\right) (-C) \mu -D^2 \left(3 e^4+24 e^2+8\right)\right)}{12 a^5 \left(1-e^2\right)^{7/2} \mu } \nonumber\\
&&-\frac{(-C)^{5/2} \left(8 a^2 h_1 C^2 \mu ^2- 3 a D^2 h_2 (-C) \mu +3 D^4 h_3\right)}{120 a^{11/2} \left(1-e^2\right)^4 \mu ^{3/2} \left(D^2+a \left(1-e^2\right) (-C) \mu \right)^{3/2}},
\n
and
\m
\label{dLtotal}
\left\langle\frac{dL}{dt}\right\rangle &=& \left\langle\frac{dL_{EM}}{dt}\right\rangle+\left\langle\frac{dL_{GW}}{dt}\right\rangle \nonumber\\
&=& -\frac{((\Delta \sigma_q)^{2}+(\Delta \sigma_g)^{2}) (-C) \left(D^2 \left(e^2+2\right)+2 a \left(1-e^2\right) (-C) \mu \right)}{3 a^3 \left(1-e^2\right)^{3/2} \sqrt{ \left(D^2+a \left(1-e^2\right) (-C) \mu \right)}} \nonumber\\
&&-\frac{(-C) \left(16 a^2  h_4 C^2 \mu ^2 - a D^2 h_5 (-C) \mu +D^4h_6 \right)}{20 a^5 \left(1-e^2\right)^{7/2} \mu ^2 \sqrt{ \left(a \left(1-e^2\right) (-C) \mu +D^2\right)}},
\n
where
\ee
h_1 = \left(1-e^2\right)^2 (37 e^4+292 e^2+96), \quad h_2 = 45 e^8+1005 e^6+670 e^4-1448 e^2-272, \quad h_3 = 5 e^6+90 e^4+120 e^2+16,
\q
and
\ee
h_4 = \left(1-e^2\right)^2 (7 e^2+8), \quad
h_5 = 31 e^6+297 e^4-192 e^2-136, \quad
h_6 = 3 (e^2+8) e^2+8.
\q
Here, we have used the relation of $a$, $e$ and $\theta$:
\m
\label{Relation}
\tan^2 (\theta)=\frac{\mu(-C)a(1-e^2)}{D^2}.
\n
The gravitational and electromagnetic radiations do not change the transverse components $L^{i}$ for $i = 1, 2$ and preserve the direction since $\left\langle\dot{L}_{GW}^i\right\rangle = \left\langle\dot{L}_{EM}^1\right\rangle = 0$ and while $\dot{L}^3 \neq 0$ implies that the magnitude of $\bm{L}$ decreases.
In principle, we can also rewrite~\eqref{dEtotal} and \eqref{dLtotal} as functions of $a$ and $\theta$ or functions of $e$ and $\theta$.
Eqs.~\eqref{dEtotal} and \eqref{dLtotal} as functions of Keplerian parameters $a$ and $e$ recover the results for the binary of charge-neutral black holes in~\cite{Liu:2020cds,Peters:1964zz}.

\section{\label{Evolution} Evolution of orbits}

The explicit formulae for emissions of energy and angular momentum due to gravitational and electromagnetic radiations enable us to calculate the evolution of orbits through two Keplerian parameters $a$, $e$ and another parameter $\theta$ due to the presence of a magnetic charge.
In doing so, we note that the emissions of energy ~\eqref{dEtotal} and angular momentum~\eqref{dLtotal} depends on three parameters $a, e,$ and $\theta$ and that $E = C/2a$ and $L = \sqrt{{a \left(1-e^2\right) (-C) \mu }+D^2}$. Applying the chain rule for differentiation, we have
\m
\frac{dE}{da}\frac{da}{dt}=\left\langle\frac{dE}{dt}\right\rangle, \quad
\frac{dL}{da}\frac{da}{dt}+\frac{dL}{de}\frac{de}{dt}=\left\langle\frac{dL}{dt}\right\rangle.
\n
Further, we separately consider contributions from the electromagnetic radiation and from the gravitational  radiation to the rates of the semimajor axis and eccentricity:
\m
\label{dadt}
\frac{da}{dt}=\frac{da_{EM}}{dt}+\frac{da_{GW}}{dt},
\n
\m
\label{dedt}
\frac{de}{dt}=\frac{de_{EM}}{dt}+\frac{de_{GW}}{dt},
\n
where $da_{EM}/dt$, $da_{GW}/dt$ and $de_{EM}/dt$ and $de_{GW}/dt$ follow from the relations
\m
\frac{dE}{da}\frac{da_{EM}}{dt}=\left\langle\frac{dE_{EM}}{dt}\right\rangle,
\quad
\frac{dL}{da}\frac{da_{EM}}{dt}+\frac{dL}{de}\frac{de_{EM}}{dt}=\left\langle\frac{dL_{EM}}{dt}\right\rangle,
\n
and
\m
\frac{dE}{da}\frac{da_{GW}}{dt}=\left\langle\frac{dE_{GW}}{dt}\right\rangle,
\quad
\frac{dL}{da}\frac{da_{GW}}{dt}+\frac{dL}{de}\frac{de_{GW}}{dt}=\left\langle\frac{dL_{GW}}{dt}\right\rangle,
\n
Finally, we obtain
\begin{widetext}
{
\ee
\label{daGW}
\a
\frac{da_{GW}}{dt} = \frac{(-C)^{3/2} \left(-8 a^2 \left(e^2-1\right)^2 h_1 C^2 \mu ^2+3 a D^2 h_2 (-C) \mu -3 D^4 h_3\right)}{60 a^{7/2} \left(e^2-1\right)^4 \left(\mu  \left(D^2-a \left(e^2-1\right) (-C) \mu \right)\right)^{3/2}},
\b
\q
\ee
\label{deGW}
\begin{aligned}
\frac{de_{GW}}{dt}=-\frac{e \Bigl(8 a^2 h_7 C^2 \mu ^2 - 3 a D^2 h_8 (-C) \mu +33 D^4 h_9 \Bigr)}{120 a^6 \left(1-e^2\right)^{9/2} \mu ^3},
\end{aligned}
\q
\ee
\label{daEM}
\a
\frac{da_{EM}}{dt}=-\frac{(\Delta \sigma_q)^{2}+(\Delta \sigma_g)^{2}) \Bigl(D^2 \left(3 e^4+24 e^2+8\right)-4 a \left(e^4+e^2-2\right) (-C) \mu \Bigr)}{6 a^3 \left(1-e^2\right)^{7/2} \mu },
\b
\q
\ee
\label{deEM}
\a
\frac{de_{EM}}{dt}=-\frac{((\Delta \sigma_q)^{2}+(\Delta \sigma_g)^{2}) e \left(7 D^2 \left(e^2+4\right)+12 a \left(1-e^2\right) (-C) \mu \right)}{12 a^4 \left(1-e^2\right)^{5/2} \mu },
\b
\q
}
\end{widetext}
where
\ee
h_7 = \left(1-e^2\right)^2 \left(121 e^2+304\right), \quad
h_8 = 107 e^6+1537 e^4-308 e^2-1336, \quad
h_9 = e^4+12 e^2+8.
\q
Noting again that $0\leq e <1$ and $C<0$ for arbitrary $q_1, q_2, g_1, g_2, m_1$ and $m_2$, we always have $da_{GW}/dt<0$, $de_{GW}/dt \leq 0 $ and $da_{EM}/dt <0$ and $de_{EM}/dt \leq 0$. Thus, we conclude that the semimajor axis and the eccentricity decreases, that is, $da/dt <0$ and $de/dt \leq 0$. In the particular case of an initial $e=0$ and a circular orbit, the eccentricity remains zero
\ee
\a
\frac{de_{GW}}{dt}=\frac{de_{EM}}{dt}=0.
\b
\q
while the semimajor axis decreases as
\ee
\a
\frac{da_{GW}}{dt}=-\frac{4 \left(a (-C) \mu +D^2\right) \left(16 a (-C) \mu +D^2\right)}{5 a^5 \mu ^3},
\b
\q
\ee
\a
\frac{da_{EM}}{dt}=-\frac{4 ((\Delta \sigma_q)^{2}+(\Delta \sigma_g)^{2}) \left(a (-C) \mu +D^2\right)}{3 a^3 \mu }.
\b
\q

A few comments are in order.
\begin{itemize}
\item
The purely electric or magnetic charges or $q_2/q_1= g_2/g_1$ correspond to $D=0$, in which $\bm{\tilde{L}}=\bm{L}$ and the binary follows a Keplerian orbit in a fixed plane~\cite{Liu:2020cds,Liu:2020vsy}. The emission of gravitational and electromagnetic radiations does not change the orbital plane but the semimajor axis and the eccentricity decreases.

\item
When $D \neq 0$, using $L=|D|/\cos \theta$ and $ d\theta /dt =\left\langle dL/dt \right\rangle / (dL/d\theta)$, the conic angle for the Poincare cone changes as
\m
\frac{d\theta}{dt}=\left\langle\frac{dL}{dt}\right\rangle \frac{\cos (\theta ) \cot (\theta )}{|D|}.
\n
The angle decreases, $d\theta / dt <0$ since $\left\langle dL/dt\right\rangle < 0$. According to \eqref{Relation}, when the semimajor axis $a$ shrinks to nearly zero, the conic angle $\theta$ also decreases to nearly zero.

\item In all cases with electric and magnetic charges, the semimajor axis $a$ always decreases during the inspiral motion due to the energy loss via gravitational and electromagnetic radiations. A circular orbit remains circular while an elliptical orbit becomes quasi-circular. 
\end{itemize}

\section{Summary and Discussion}

In this paper, we have reviewed the orbital motion of a binary of black holes with electric and/or magnetic charges and found the loss rates of energy and angular momentum due to the gravitational and electromagnetic radiations. Then we have calculated the evolution of orbits due to the loss of energy and angular momentum. Further, we have compared the evolution of the orbits for binary of black holes with or without electric and magnetic charges in Table \ref{table1}.

A binary of black holes without any charge or with pure electric charges follows the Keplerian orbit: circular and elliptical one on a fixed plane. On the other hand, a binary both with electric and magnetic charges, that is, the binary of dyonic black holes has an additional velocity-dependent force, which results in a generalized angular momentum with respect to which the orbital angular momentum precesses. Thus the binary of dyonic black holes is a Keplerian-type orbit on a Poincare cone with finite or infinite winding numbers. The orbit with an infinite winding number exhibits chaotic behavior.


\begin{table}
\renewcommand{\arraystretch}{1.5}
\begin{center}
\begin{tabular}{ |p{4cm}|m{1cm}|m{7.5cm}| }
\hline
 & \large{Orbits} & \large The evolution of the orbits   \\
\hline
\large Schwarzschild binary black holes \cite{Peters:1963ux,Peters:1964zz} & 2D & $\frac{da}{dt}=-\frac{2 \left(37 e^4+292 e^2+96\right) \mu  M^2}{15 a^3 \left(1-e^2\right)^{7/2}},\frac{de}{dt}=-\frac{e \left(121 e^2+304\right) \mu  M^2}{15 a^4 \left(1-e^2\right)^{5/2}}$ \\ \hline
 \large \multirow{2}{4cm}{Binary black holes with electric charges\cite{Liu:2020cds}} & \multirow{2}{1cm}{2D} & $\frac{da}{dt}=-\frac{2 \left(37 e^4+292 e^2+96\right) C^2}{15 a^3 \left(1-e^2\right)^{7/2} \mu }-\frac{2 (\Delta \sigma_q)^{2} (-C)}{3 a^2 \left(1-e^2\right)^{5/2}},$ \\
&  &  $\frac{de}{dt}=-\frac{e \left(121 e^2+304\right) C^2}{15 a^4 \left(1-e^2\right)^{5/2} \mu }-\frac{(\Delta \sigma_q)^{2} e (-C)}{a^3 \left(1-e^2\right)^{3/2}}$  \\
\hline
\large {\multirow{4}{4cm}{Binary black holes with electric and magnetic charges\cite{Liu:2020vsy,Liu:2020bag}}}&  & $\frac{da}{dt}=\frac{(-C)^{3/2} \left(-8 a^2 \left(e^2-1\right)^2 h_1 C^2 \mu ^2+3 a D^2 h_2 (-C) \mu -3 D^4 h_3\right)}{60 a^{7/2} \left(e^2-1\right)^4 \left(\mu  \left(D^2-a \left(e^2-1\right) (-C) \mu \right)\right)^{3/2}}$ \\
 & \multirow{1}{*}{3D} &  $-\frac{(\Delta \sigma_q)^{2}+(\Delta \sigma_g)^{2}) \Bigl(D^2 \left(3 e^4+24 e^2+8\right)-4 a \left(e^4+e^2-2\right) (-C) \mu \Bigr)}{6 a^3 \left(1-e^2\right)^{7/2} \mu },$   \\
 &  &  $\frac{de}{dt}=-\frac{e \Bigl(8 a^2 h_4 C^2 \mu ^2 - 3 a D^2 h_5 (-C) \mu +33 D^4 h_6 \Bigr)}{120 a^6 \left(1-e^2\right)^{9/2} \mu ^3}$  \\ &  & $-\frac{((\Delta \sigma_q)^{2}+(\Delta \sigma_g)^{2}) e \left(7 D^2 \left(e^2+4\right)+12 a \left(1-e^2\right) (-C) \mu \right)}{12 a^4 \left(1-e^2\right)^{5/2} \mu }$    \\
\hline
\end{tabular}
\end{center}
\caption{\label{table1}The comparison of the evolution of the orbits for binary of black holes with or without electric and magnetic charges.  }
\end{table}

We have found the following physical interpretations for the binary of black holes with electric and/or magnetic charges.
\begin{itemize}
\item Each black hole of the binary does not lose the mass and electric and/or magnetic charges during the inspiral motion. It is the gravitational and electrostatic binding energy and the orbital angular momentum that supplies the energy and angular momentum for the gravitational and electromagnetic radiations.
However, when the binary merges, part of the mass and charges are converted to the gravitational and electromagnetic radiations and the annihilation of opposite charges adds up the burst of energy.

\item Black holes can carry electric and/or magnetic charges of the same sign or opposite signs. A binary of black holes with charges of the same sign can be formed as afar as the gravitational attraction dominates the electrostatic repulsion between two black holes such that $C < 0$ in~(\ref{C-D}). The charge of each black hole emits synclotron radiation from the circular or elliptical orbit and the electric and magnetic dipole radiations are the leading electromagnetic radiation. The electric dipole ${\bf d} = \sum_{i = 1,2} q_i {\bf r}_i = \mu \Delta \sigma_q {\bf r}$ and the magnetic dipole ${\bf m} = \sum_{i = 1,2} g_i {\bf r}_i = \mu \Delta \sigma_g {\bf r}$ give the dipole radiation. When the ratios of electric and magnetic charges to the masses are the same, the dipole radiations vanish.
    In general, a closed system of charges with the same charge to mass ratio has the electric and magnetic dipoles proportional to the center of mass, which undergoes a uniform motion and thus prohibits the dipole radiation.

\item Though the binary follows a Keplerian orbit for purely electrically or magnetically charged black holes, the gravitational radiation is in fact the post-Newtonian result of order 1.5. Similarly, electromagnetic radiation is a relativistic effect of the Maxwell theory, regardless of whether the charges move relativistically or nonrelativistically. The binary of dyonic black holes follows a Keplerian-type orbit on a Poincare cone and emits the gravitational and electromagnetic radiations similarly to the binary of black holes with electric charges. 
\end{itemize}

\begin{acknowledgments}
The authors are benefitted from helpful discussions during the 17th Italian-Korean Symposium on Relativistic Astrophysics. This work was supported by the National Research Foundation of Korea (NRF) funded by the Ministry of Education (2019R1I1A3A01063183).
\end{acknowledgments}

\bibliography{Ref}

\end{document}